\documentclass[doublecol, amssymb, amsmath]{epl2}
\usepackage{graphicx}
\usepackage{amssymb}
\usepackage{amsmath}
\usepackage[english]{babel}
\begin{document}
\title{Spin excitations in a monolayer scanned by a
  magnetic tip}
\author{M.P. Magiera \inst{1} \and L. Brendel \inst{1} \and D.E. Wolf
  \inst{1} \and U. Nowak \inst{2}} 
\institute{
   \inst{1} Department of Physics and CeNIDE, University of Duisburg-Essen,
   D-47048 Duisburg, Germany, EU \\
   \inst{2} Theoretical Physics, University of Konstanz, D-78457
   Konstanz, Germany, EU 
}


\abstract{Energy dissipation via spin excitations is investigated for
  a hard ferromagnetic tip scanning a soft magnetic monolayer. We use
  the classical Heisenberg model with Landau-Lifshitz-Gilbert (LLG) dynamics
  including a stochastic field representing finite temperatures. The
  friction force depends linearly on the velocity (provided it is
  small enough) for all temperatures. For low temperatures, the
  corresponding friction coefficient is proportional to the
  phenomenological damping constant of the LLG equation. This
  dependence is lost at high temperatures, where the friction
  coefficient decreases exponentially. These findings can be explained
  by properties of the spin polarisation cloud dragged along with the tip. 
}

\maketitle

\section{Introduction}

While on the macroscopic scale the
phenomenology of friction is well known, several new aspects are
currently being investigated on the micron and nanometer scale
\cite{Persson1998,Urbakh2004}. During the last two decades, 
the research on
microscopic friction phenomena has advanced enormously,
thanks to the development of Atomic Force Microscopy (AFM,
\cite{Gnecco2001}), which allows to measure energy dissipation 
caused by relative motion of a tip with respect to a substrate.

Recently the contribution of magnetic degrees of freedom to energy
dissipation processes has attracted increasing interest
\cite{Acharyya95,ortin98,Corberi99,Kadau08,Fusco08}. Today, magnetic
materials can be controlled down to the nanometer scale. New
developments in the data storage industry, spintronics and quantum
computing require a better understanding of tribological phenomena in
magnetic systems. For example, the reduction of heat generation in
reading heads of hard disks which work at nanometer distances is an
important issue, as heat can cause data loss.

Magnetic Force Microscopy (MFM), where both
tip and surface are magnetic, is used to investigate
surface magnetism and to visualise domain walls. Although recent
studies have attempted to measure energy dissipation between an
oscillating tip and a magnetic sample
\cite{Gruetter1997,Heinze}, the
dependence of the friction force on the tip's sliding velocity has not
been considered yet apart from a work by C.~Fusco \textit{et al.} \cite{Fusco08}
which is extended by the present work to temperatures $T{\neq}0$. 
The relative motion of the tip with 
respect to the surface can lead to the creation of spin waves which
propagate inside the sample and dissipate energy, giving rise to
magnetic friction.

We will first present a simulation model and define magnetic friction.
The model contains classical Heisenberg spins located on a rigid
lattice which interact by exchange interaction with each
other. Analogous to the reading head of a hard disc or a MFM tip, an
external fixed magnetic moment is moved
across the substrate. Using Langevin dynamics and damping, it is
possible to simulate systems at finite temperatures. The main new
results concern the temperature dependence of magnetic friction.

\section{Simulation model and friction definition}
\label{SEC:MOD}
To simulate a solid magnetic monolayer (on a nonmagnetic substrate),
we consider a two-dimensional rigid $L_x \times L_y$ lattice of
classical normalised dipole moments (``spins'') $\mathbf S_i =
\boldsymbol \mu_i/\mu_s$, where $\mu_s$ denotes the material-dependent
magnetic saturation moment (typically a few Bohr magnetons). These
spins, located at $z=0$ and with lattice spacing $a$,
represent the magnetic moments of single atoms. They can change their
orientation but not their absolute value, so that there are two
degrees of freedom per spin.  We use open boundary
conditions. A constant point dipole $\mathbf
  S_\mathrm{tip}$ pointing in the $z$-direction and located at $z=2a$
represents the magnetic tip.  It is moved parallel to the surface with
constant velocity $\mathbf v$. 

This model has only magnetic degrees of freedom and thus
  focusses on their contributions to friction. For a real tip one
  could expect that magnetic, just like nonmagnetic
  \cite{Zwoerner1998,Gnecco00} 
  interactions might also lead to atomic 
  stick-slip behaviour, and hence to phononic dissipation with a
  velocity-independent friction contribution as described by the
  Prandtl-Tomlinson model \cite{Prandtl1928,Tomlinson1929}.  
  However, this requires a periodic potential between tip and
  substrate, that is strong enough compared to the elastic deformation
  energy to allow for multiple local potential energy minima. The
  magnetic tip-substrate interactions are unlikely to be strong enough.
  
The Hamiltionan consists of two parts:
\begin{equation}
\label{eq:H}
\mathcal H = \mathcal H_\mathrm{sub} + \mathcal H_\mathrm{sub-tip}.
\end{equation}
The first one represents the internal
ferromagnetic short-range interaction within the substrate. The second
one describes the long-range coupling between the substrate and the tip.

The interaction between the substrate moments is modeled by the
anisotropic classical Heisenberg model,
\begin{equation}
  \label{eq:H_sub}
  \mathcal H_\mathrm{sub} = -J\sum_{\langle i,j\rangle} \mathbf S_i \cdot \mathbf S_j - d_\mathrm{z} \sum_\mathrm{i=1}^N{S_{i,z}^2}.
\end{equation}
$J>0$ describes the ferromagnetic exchange interaction between two
nearest neighbours, expressed by the angular brackets $\langle i,j\rangle$.
$d_\mathrm{z}<0$ quantifies the anisotropy, which prefers in-plane
orientations of the spins. The
dipole-dipole-interaction between the substrate spins is neglected, because
it is much weaker than the exchange interaction.
A quantitative comparison of our simulation results with the ones
obtained in \cite{Fusco08}, where the dipole-dipole-interaction inside the
substrate was taken into account, justifies this approximation, 
which reduces simulation time enormously. 

The
long-range interaction between substrate and tip is described by a
dipole-dipole interaction term
\begin{equation}
  \label{eq:H_sub-tip}
  \mathcal H_\mathrm{sub-tip} = - w \sum_{i=1}^N{\frac{3
  ~(\mathbf S_i \cdot \mathbf e_i) (\mathbf S_\mathrm{tip} \cdot
  \mathbf e_i) - \mathbf S_i \cdot \mathbf S_\mathrm{tip} } 
  {R_i^3}}, 
\end{equation}
where $R_i = \left |\mathbf R_i \right |$ denotes the norm of the
distance vector $\mathbf R_i  = \mathbf r_i - \mathbf r_\mathrm{tip}$,
and $\mathbf e_i$ its unit vector $\mathbf e_i = \mathbf 
R_i / R_i $. $\mathbf r_i$ and $\mathbf r_\mathrm{tip}$ denote the
position vectors of the substrate spins and the tip respectively.
$w$ quantifies the dipole-dipole coupling of the substrate and
the tip. Note, however, that the results of the present study only
depend on the combination $w |\mathbf S_\mathrm{tip}|$, which is the
true control parameter for the tip-substrate coupling.

The proper equation of motion of the magnetic moments is the
\textit{Landau-Lifshitz-Gilbert}
(LLG,\cite{LandauLifshitz1935}) equation,
\begin{equation}
  \label{eq:LLG}
  \frac{\partial}{\partial t} \mathbf S_i = -\frac{\gamma}{(1 +
  \alpha^2) \mu_s} \left[\mathbf S_i \times \mathbf h_i + \alpha ~
  \mathbf S_i \times(\mathbf S_i \times \mathbf h_i) \right], 
\end{equation}
which is equivalent to the Gilbert equation of motion
\cite{Gilbert1955}:
\begin{equation}
\label{eq:ge}
\frac{\partial}{\partial t} \mathbf S_i = - \frac{\gamma}{\mu_s}
\mathbf S_i \times \left [ \mathbf h_i - \frac{\alpha \mu_s}{\gamma}
  \frac{\partial \mathbf S_i}{\partial t} \right ]. 
\end{equation}
The first term on the right-hand side of eq. (\ref{eq:LLG}) describes
the dissipationless precession of each spin in the effective field $\mathbf h_i$ (to be specified below).  The second
term describes the relaxation of the spin 
towards the direction of $\mathbf h_i$.
$\gamma$ denotes the gyromagnetic ratio (for free 
electrons $\gamma = 1.76086 \times 10^{11} {\rm s}^{-1} {\rm T}^{-1}$),
and $\alpha$ is a phenomenological, dimensionless damping parameter.

The effective field contains contributions from the tip and from
the exchange interaction, as well as a thermally fluctuating term
$\boldsymbol\zeta_i$\cite{Neel1949,Brown1963}, 
\begin{equation}
  \label{eq:h_i}
  \mathbf h_i = -\frac{\partial \mathcal H}{\partial \mathbf S_i} + \boldsymbol \zeta_i(t).
\end{equation}
The stochastic, local and time-dependent vector $ \boldsymbol
\zeta_i(t)$ expresses a ``Brownian rotation'', which is caused by the
heat-bath connected to each magnetic moment. In our simulations this
vector is realised by uncorrelated random
numbers with a Gaussian distribution, which satisfy the relations
\begin{eqnarray}
\label{eq:noise_1}
\langle  \boldsymbol \zeta_i (t) \rangle &=& 0 \;\; \mbox{   and}\\
\label{eq:noise_2}
\langle  \zeta_i^\kappa (t) \zeta_j^\lambda (t') \rangle &=& 
2 \frac{\alpha \mu_s}{\gamma}\ k_BT\delta_{i,j}\delta_{\kappa,\lambda} \delta (t - t'),
\end{eqnarray}
where $T$ is the temperature, $\delta_{i,j}$
expresses that the noise at different lattice sites is uncorrelated, and
$\delta_{\kappa, \lambda}$ refers to the absence of correlations among different
coordinates.

To find a quantity which expresses the friction occurring in the
system, it is helpful to discuss energy transfers between tip,
substrate and heat-bath first. It
is straightforward to separate the time derivative of the
system energy, eq. (\ref{eq:H}), into an explicit and an implicit one,
\begin{equation}
  \label{eq:P_tot}
  \frac{d\mathcal H}{dt} = \frac{\partial \mathcal H}{\partial t} +
  \sum_{i=1}^N {\frac{\partial \mathcal H}{\partial \mathbf S_i}\cdot
  \frac{\partial \mathbf S_i}{\partial t}}. 
\end{equation}
The explicit time dependence is exclusively due to the tip motion. The
energy transfer between the tip and the substrate is expressed by
the first term of eq.~(\ref{eq:P_tot}), which justifies to call it
the ``pumping power'' $\mathcal P_\mathrm{pump}$:
\begin{align}
  \label{eq:P_pump}
  & \mathcal P_\mathrm{pump} = \frac{\partial \mathcal H}{\partial t} 
= \frac{\partial \mathcal H_\mathrm{sub-tip}}{\partial \mathbf
  r_\mathrm{tip}} \cdot \mathbf v  \nonumber 
\\  = & w \sum_\alpha v_\alpha \sum_{i=1}^{N} \frac{3}{R_i^4} 
\bigg \{ (\mathbf S_i \cdot \mathbf e_i ~ e_{i,\alpha} - S_{i,
  \alpha}) (\mathbf S_\mathrm{tip} \cdot \mathbf e_i)   
\\ & + (\mathbf S_\mathrm{tip} \cdot \mathbf e_i~e_{i,\alpha} -
S_\mathrm{tip, \alpha}) (\mathbf S_i \cdot \mathbf e_i) \nonumber 
\\ &- \mathbf S_i \cdot \mathbf S_\mathrm{tip} ~e_{i,\alpha} \nonumber  +
3e_{i,\alpha} (\mathbf S_{i} \cdot \mathbf e_i)(\mathbf S_\mathrm{tip} \cdot
\mathbf e_i) \bigg \} \nonumber 
\end{align}

At any instance, the substrate exerts a force $-\frac{\partial
  \mathcal H_\mathrm{sub-tip}}{\partial \mathbf r_\mathrm{tip}}$ on
  the tip. Due to Newton's third law, $\mathcal P_\mathrm{pump}$ is
  the work per unit time done by the tip on the substrate. Its 
time and thermal average
$\langle\mathcal P_\mathrm{pump}\rangle$ is the average rate at which 
energy is pumped into the spin system. In a steady
state it must be equal to the average dissipation rate, \textit{i.e.} to the net energy
transferred to the heat bath per unit time due to spin relaxation.
The magnetic friction force can therefore be calculated by
\begin{equation}
  \label{eq:F}
  F = \frac{\left <\mathcal P_\mathrm{pump} \right >}{v}.
\end{equation}

The second term of eq.~(\ref{eq:P_tot}) describes the energy transfer
between the spin system and the heat bath. 
Inserting eq. (\ref{eq:ge}) into 
$\mathcal P_\mathrm{diss}=-\sum_{i=1}^N {\frac{\partial \mathcal
    H}{\partial \mathbf S_i}\cdot \frac{\partial \mathbf S_i}{\partial
    t}}$ leads to 
\begin{eqnarray}
\label{eq:P_diss}
\mathcal P_\mathrm{diss} &=& \sum_{i=1}^N
\frac{\partial \mathcal H}{\partial \mathbf S_i} \cdot 
  \left [ \frac{\gamma}{\mu_s}
  \mathbf S_i \times \boldsymbol \zeta_i - \alpha \mathbf S_i \times
  \frac{\partial \mathbf S_i}{\partial t} \right ] \nonumber \\  
&=& - \mathcal P_\mathrm{therm} + \mathcal P_\mathrm{relax}.
\end{eqnarray}
The first term, $\mathcal P_\mathrm{therm}$ containing $\zeta_i$,
describes, how much energy is transferred into the spin system due to
the thermal perturbation by the heat bath. The second term, $\mathcal
P_\mathrm{relax}$ proportional to the damping constant $\alpha$,
describes the rate of energy transfer into the heat bath due to the
relaxation of the spins. 

At $T=0$, $\mathcal P_\mathrm{therm}$ is zero. The spins are only perturbed
by the external pumping at $v\neq 0$. Then 
\begin{align}
\label{eq:P_diss2}
\mathcal P_\mathrm{relax}&=\mathcal P_\mathrm{diss} 
  = \frac{\gamma
  \alpha}{\mu_s(1+\alpha^2)}\sum_{i=1}^N\left(\mathbf S_i \times
 \mathbf h_i \right)^2,
\quad (T=0),
\end{align}
where for the last transformation we used eq.~(\ref{eq:LLG}) in order
to show explicitly the relationship between dissipation rate and
misalignment between spins and local fields.

\begin{figure}[]
\centering
\includegraphics[width=0.4\textwidth]{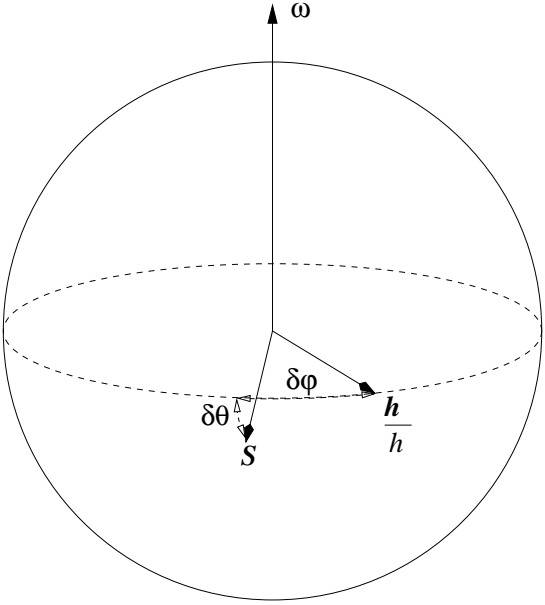}
\caption{\label{fig:retardation}A single spin in a magnetic field
  rotating with angular velocity $\boldsymbol{\omega}$, is dragged
  along with a phase shift $\delta\varphi$ and aquires an out of plane
  component $\delta\theta$.}  
\end{figure}

It will be instructive to compare the magnetic substrate scanned by a
dipolar tip with a much simpler system, in which the substrate is
replaced by a single spin $\mathbf S$ subjected to an external field
$\mathbf h(t)$ that rotates in the plane perpendicular to a constant
angular velocity $\boldsymbol{\omega}$ (replacing the tip velocity).  It is
straight forward to obtain the steady state solution for $T=0$, where
in the co-rotating frame $\mathbf S$ is at rest. 
$\mathbf S$ lags behind $\mathbf h/h$ by an angle $\delta\varphi$ and gets a
component $\delta\theta$ in $\boldsymbol\omega$-direction
(cf. fig.~\ref{fig:retardation}), which are 
in first order given by 
\begin{equation}
  \label{eq:adiabat}
\frac{\delta\varphi}{\alpha} = \delta\theta = \frac{\omega\,\mu_s}{h\,\gamma} + \mathcal\!O\left(\!\left(\frac{\omega\,\mu_s}{h\,\gamma}\right)^3\right).
\end{equation}
Inserting this into the ($N{=}1$)-case of eq.~(\ref{eq:P_diss2}) yields a
dissipation rate of $\mathcal
P_\mathrm{diss}=\alpha\omega^2\mu_s/\gamma$, which corresponds to a
``viscous'' friction $F=\mathcal
P_\mathrm{diss}/\omega \propto \alpha \omega$.

It is instructive to give a simple physical explanation for
eq.~(\ref{eq:adiabat}), instead of presenting the general solution, which
can be found in \cite{Magiera_Diplom}.  
Two timescales exist in the system, which can be
readily obtained from eq.~(\ref{eq:LLG}); first, the inverse Lamor frequency
$\tau_\mathrm{Lamor}=(1+\alpha^2)\mu_s/\gamma h$, and second, the
relaxation time
$\tau_\mathrm{relax} {=} \tau_\mathrm{Lamor}/\alpha$. 
They govern the time evolution of $\delta\varphi$ and $\delta\theta$.
In leading order,
\begin{eqnarray}
\delta \dot \theta &=& \frac{\delta\varphi}{\tau_\mathrm{Lamor}} 
- \frac{\delta\theta}{\tau_\mathrm{relax}},\\
\delta \dot \varphi &=& \omega - \frac{\delta\theta}{\tau_\mathrm{Lamor}}
- \frac{\delta\varphi}{\tau_\mathrm{relax}}.
\end{eqnarray}
The first equation describes, how $\delta\theta$ would increase by
precession of the spin around the direction of the field, which is
counteracted by relaxation back towards the equator. The second
equation describes that without relaxation into the field direction,
$\delta\varphi$ would increase with velocity $\omega$ minus the
azimuthal component of the precession velocity, which is in leading
order proportional to $\delta\theta$. Setting the left hand sides to
zero in the steady state, immediately gives the solution (\ref{eq:adiabat}).

In the ($T=0$)-study \cite{Fusco08}, the time average $\langle \mathcal
P_\mathrm{relax} \rangle/v $ was used to calculate the friction
force. As pointed out above, this quantity agrees with
(\ref{eq:F}) in the steady state. For finite temperatures,
however, (\ref{eq:F}) is numerically better behaved than 
$\langle \mathcal P_\mathrm{relax} \rangle/v $. The reason is the following:

For $T\neq 0$, the spins are also thermally agitated, even
without external pumping, when the dissipation rate $\mathcal P_\mathrm{diss}$
vanishes. This
shows that the two terms $\mathcal P_\mathrm{therm}$ and $\mathcal
P_\mathrm{relax}$ largely compensate each other, and only their
difference is the dissipation rate we are interested in. This fact
makes it difficult to evaluate (\ref{eq:P_diss}) and is the reason why
we prefer to work with (\ref{eq:F}) as the definition of the friction
force. 

We have also analyzed the fluctuations of the friction force
(\ref{eq:F}). The power spectrum has a distinct peak at frequency
$v/a$, which means that the dominant temporal fluctuations are due to
the lattice periodicity with lattice constant $a$. A more complete
investigation of the fluctuations, which should also take into
account, how thermal positional fluctuations influence the friction
force, remains to be done.

\section{Technical remarks}

Because of the vector pro\-duct in eq.~(\ref{eq:LLG}), the noise
$\boldsymbol \zeta_i$ enters in a multiplicative way, calling for
special attention to the interpretation of this stochastic
differential equation (Stratonovich vs. It\^o sense). The
physical origin of the noise renders it generically \textit{coloured}
and thus selects the Stratonovich interpretation as the appropriate
one (Wong-Zakai theorem \cite{Horsthemke1983}), in which its
appearance as white noise is an idealisation. 
Accordingly we employ the Heun integration scheme \cite{Garcia1998}.
After each time step the spins are rescaled so that their length
remains unchanged.

To get meaningful results, it is of prime importance to reach a 
steady state. The initial configuration turned out to be a crucial
factor for achieving this within acceptable computing time. Therefore
a long initialisation run is performed, before the tip motion starts.

Moreover, the system size is another limiting factor. In order to
avoid that the tip reaches the system boundary before the steady state is
reached, we use a ``conveyor belt method'' allowing to do the simulation 
in the comoving frame of the tip. 
The tip is placed in a central point above the substrate
plane, \textit{e.g.} at $((L_x+1)/2, (L_y+1)/2)$. After an equilibration
time, the tip starts to move with fixed velocity in
$x$-direction. When it passed exactly 
one lattice constant $a$, the front line (at $x=L_x$) is
duplicated and added to the lattice at $x=L_x+1$. Simultaneously,
the line at the opposite boundary of the system (at $x=1$) is deleted,
so that the simulation cell is of fixed size and consists of the 
$L_x \times L_y$ spins centered around the tip position, with open boundaries.
Note that this is different from periodic boundary conditions, because the spin configuration deleted at one side is different from the one added at the opposite side.
We compared the results obtained for a small system in the
co-moving frame of the tip with some runs for a system that was long enough
in $x$-direction that the steady state could be reached in the rest frame
of the sample, and confirmed that the same friction results and steady
state properties could be obtained with drastically reduced
computation time.

It is convenient to
rewrite the equations of motion in natural units. An energy unit is
prescribed by the exchange energy $J$ of two magnetic moments. It
rescales the energy related parameters $d_z$ and $w$ as well as
the simulated temperature,
\begin{equation}
k_B T' = \frac{k_B T}{J}.
\end{equation}
The rescaled time further depends on the material constants $\mu_s$
and $\gamma$, 
\begin{equation}
t' = \frac{J \gamma}{\mu_s} ~t.
\end{equation}
A length scale is given by the lattice constant $a$, so a natural
velocity for the system can be defined, 
\begin{equation}
v' = \frac{\mu_s}{\gamma J a}~ v.
\end{equation}
From the natural length and energy, the natural force results to:
\begin{equation}
F' = \frac{a}{J} F
\end{equation}
From now on, all variables are given in natural units, and we will drop
the primes for simplicity.
The typical extension of the simulated lattices is $50\times30$, which
was checked to be big enough to exclude finite size-effects. For the
tip coupling, we chose large values (\textit{e.g.} $w \mathbf S_\mathrm{tip} = (0,
0, -10)$), to get a large effective field on the substrate. Usually it
is assumed that the dipole-dipole coupling constant has a value of
about $w=0.01$, which means that the magnetic moment of the tip is
chosen a factor of 1000 times larger than the individual substrate
moments.  The anisotropy constant is set to $d_z=-0.1$ in all
simulations. The damping constant $\alpha$ is varied from $0.1$ to the
quite large value $1$. At finite temperatures typically 50 simulation runs
with different random number seeds are performed to get reliable
ensemble averages.

\section{Simulation results}
In \cite{Fusco08} it was found that the magnetic friction force
depends linearly on the scanning velocity $v$ and the
damping constant $\alpha$ 
for small velocities ($v{\le}0.3$).
For higher velocities the friction force
reaches a maximum and then decreases. 
In this work we
focus on the low-velocity regime with the intention to shed more light
on the friction mechanism and its temperature dependence.\footnote{It should be noted that the smallest tip velocities we
  simulated, are of the order of $10^{-2} (aJ\gamma/\mu_s)$, which is
  still fast compared to typical velocities in friction force
  microscopy experiments.}

\begin{figure}[bt]
\includegraphics[width=0.49\textwidth]{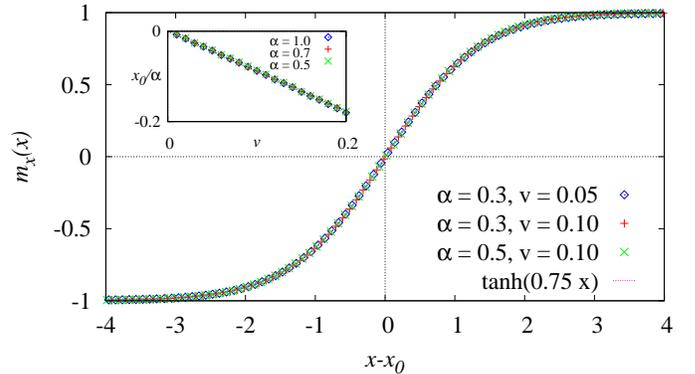}
\caption{\label{fig:tanh} Local magnetisation ($x$-component) at $T{=}0$
along the lattice axes in $x$-direction which are closest to the tip (\textit{i.e.} at 
  $y{=}{\pm}0.5$). Analogous to a domain wall one finds a
  $\tanh{x}$-profile. Depending on the damping constant $\alpha$ and
  the velocity $v$, its zero is shifted backwards from the tip
  position by a value $x_0{\approx}{-}0.88\alpha v$, as shown in the inset.} 
\end{figure}

\subsection{Adiabatic approximation at $T{=}0$ }
If we assume the field $\mathbf h$ for each spin to vary slowly enough
to allow the solution (\ref{eq:adiabat}) to be attained as
adiabatic approximation, the linear dependence $F\propto\alpha v$ from
\cite{Fusco08} follows immediately: At every point, the temporal
change of the direction of $\mathbf h$, defining a local
$\omega$ for (\ref{eq:adiabat}), is proportional to $v$.
\begin{figure*}[]
\centering
\includegraphics[width=1.00\textwidth]{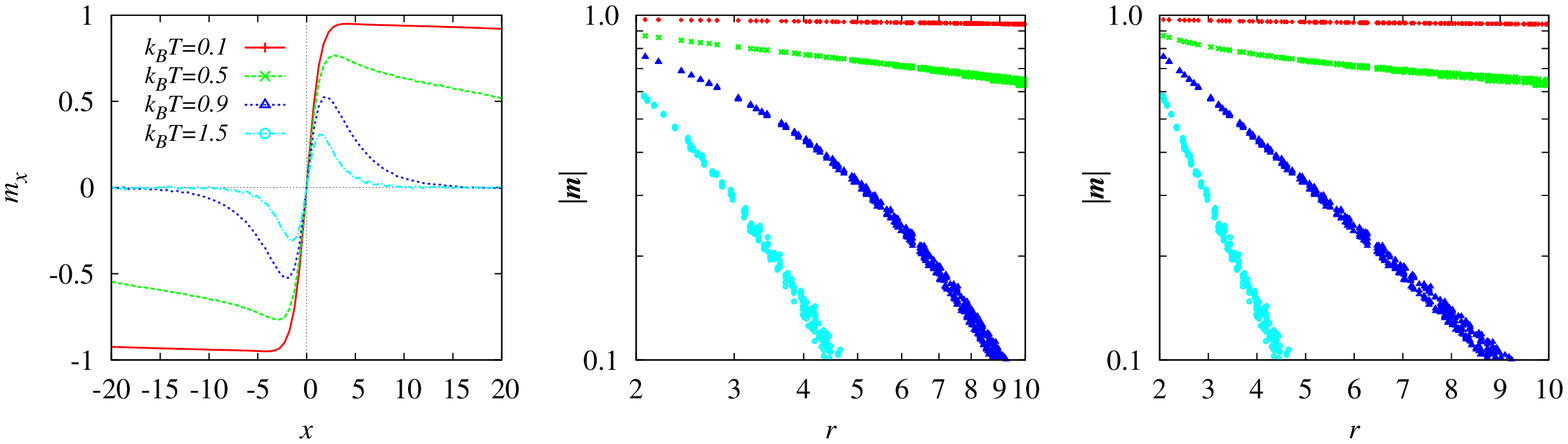}
\caption{
\label{fig:tg0_tanh}
Left:  Magnetisation profiles as in fig.~\ref{fig:tanh} for
  several temperatures with $w S_\mathrm{tip}=10$, $\alpha=0.5$ and $v=0.01$. Middle and
  right: Absolute value of the average magnetisation as a function of
  the distance $r$ from $(x,y)=(0,0)$, directly underneath the
  tip. For small temperatures (upper two curves) it decreases with a power law
  (cf. double-logarithmic plot, middle),
  for high temperatures (lower two curves) it decreases exponentially
  (cf. semi-logarithmic plot, right).} 
\end{figure*}
We confirmed the validity of the adiabatic approximation numerically
by decomposing $\mathbf S-\mathbf h/h$  with respect to the local
basis vectors 
$\partial_t (\mathbf h/h)$, $\mathbf h$, and their
cross-product, all of them appropriately normalized. In other words,
we extracted $\delta \varphi$ and $\delta \theta$ directly and 
found them in excellent agreement with (\ref{eq:adiabat}).

The lag of $\mathbf S$ with respect to $\mathbf h$ manifests itself
also macroscopically in the magnetisation field as we will show now.
The tip-dipole
is strong enough to align the substrate spins to nearly
cylindrical symmetry. Since the anisotropy is chosen to generate an
easy plane ($d_z{<}0$), spins far away from the tip try to lie in the
$xy$-plane, while close to the tip they tilt into the $z$-direction.
This is displayed in fig.~\ref{fig:tanh} where the
$x$-component of the local magnetisation is shown along a line
in $x$-direction for a fixed $y$-coordinate. Remarkably, these
magnetisation profiles for different values of $v$ and $\alpha$
collapse onto a unique curve, if they are shifted by corresponding offsets 
$x_0$ with respect to the tip position. As expected from
(\ref{eq:adiabat}), the magnetisation profile stays behind the tip by
a ($y$-dependent) shift $x_0\propto \alpha v$ (cf.\ inset of
fig.~\ref{fig:tanh}).

\subsection{Friction at $T{>}0$}
With increasing temperature the magnetisation induced by the tip
becomes smaller, as shown in fig.~\ref{fig:tg0_tanh}. One can distinguish a
low temperature regime, where the local magnetisation decreases
algebraically with the distance from the tip, and a high temperature
regime, where it decreases exponentially. The transition between these
regimes happens around $T\approx 0.7$.  

For all temperatures the friction force $F$ turns out to be
proportional to the velocity (up to $v \approx 0.3$), as for $T{=}0$,
with a temperature dependent friction coefficient $F/v$. 

\begin{figure}[]
\centering
\includegraphics[width=0.49\textwidth]{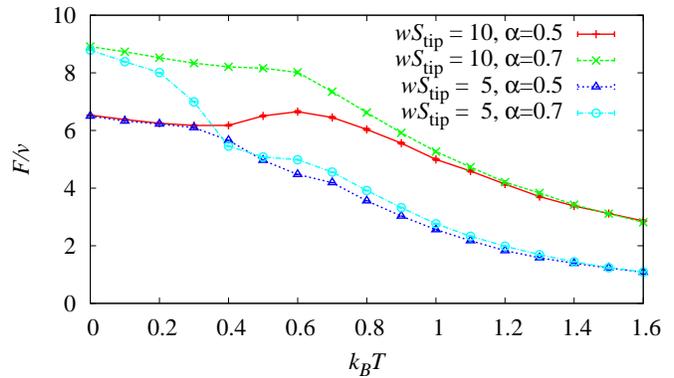}
\caption{\label{fig:tg0_coefficients}Friction coefficients for
  different $\alpha$, $wS_\mathrm{tip}$ and $k_BT$. One can distinguish
  between a low temperature regime, where the friction coefficient
  depends on $\alpha$ but not on $wS_\mathrm{tip}$, and a
  high temperature regime, where it depends on $wS_\mathrm{tip}$ but
  not on $\alpha$.} 
\end{figure}

The two temperature regimes manifest themselves also here, as shown in
fig.~\ref{fig:tg0_coefficients}: For low temperatures the friction
coefficients depend nearly linearly on $\alpha$, reflecting the
$T{=}0$ behaviour. Towards the high temperature regime, however, the
$\alpha$-dependence vanishes, and all friction coefficients merge into
a single exponentially decreasing function.

The low temperature behaviour can be understood essentially along the
lines worked out for $T{=}0$, as result of a delayed, deterministic
response (precession and relaxation) to the time dependent tip field.
At high temperatures, however, friction results from the ability of
the tip to propagate partial order through the thermally disorderd
medium. The magnetisation pattern in the wake of the tip no longer
adapts adiabatically to the dwindling influence of the tip, but decays
due to thermal disorder. Then, a rising temperature lets the ordered
area around the tip shrink which leads to the exponential decrease of
the friction coefficient. However, it increases with the tip strength,
$wS_\mathrm{tip}$, as stronger order can be temporarily forced upon
the region around the tip. By contrast, the tip strength looses its
influence on friction in the limit $T\rightarrow 0$, because the
substrate spins are maximally polarised in the tip field . 

\begin{figure}[]
\centering
\includegraphics[width=0.49\textwidth]{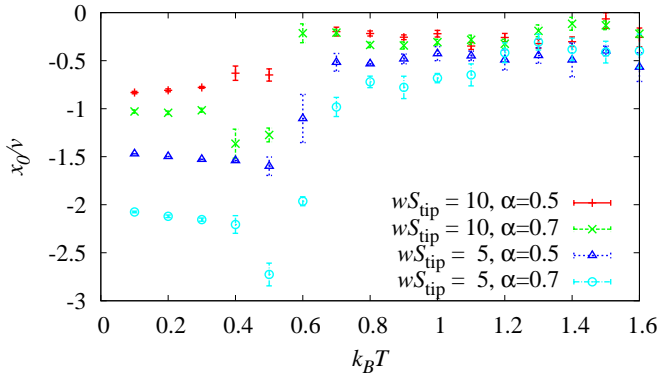}
\caption{\label{fig:ret_const}Distance $x_0$ by which the
  magnetisation pattern lags behind the tip is proportional to $v$ for
  all temperatures. The proportionality constant depends on $\alpha$
  only in the low temperature regime.} 
\end{figure}

This picture of the two temperature regimes is supported by the
distance $x_0$, by which the 
magnetisation pattern lags behind the tip. It is proportional to $v$
for all temperatures, but $\alpha$-dependent only in the low
temperature regime, cf. fig.~\ref{fig:ret_const}. 

\section{Conclusion and outlook}

In this work, we could explain the low-velocity, zero-temperature
findings from \cite{Fusco08}, namely that the magnetic friction force
in the Heisenberg model has a linear velocity dependence with a
coefficient proportional to the damping constant $\alpha$. 
In the spin polarisation cloud dragged along with the tip, each
substrate spin follows the local field with a lag proportional to the
frequency of the field change and to $\alpha$. Moreover, the
magnetisation pattern around the tip gets distorted due to
precession. These effects directly give rise to the observed magnetic
friction and could be evaluated quantitatively by means 
of a single spin model.

Second, for the first time the temperature dependence of magnetic
friction in the Heisenberg model was investigated in the framework of
Landau-Lifshitz-Gilbert (LLG) dynamics with a stochastic contribution to the
magnetic field. Two regimes were found, which can be
characterised by their different relaxation behaviour. 
While in the low-temperature regime the response of the system on the
perturbation due to the moving tip is dominated by the deterministic
precession and relaxation terms in the LLG equation, thermal
perturbations competing with the one caused by the moving tip are essential
in the high-temperature regime. This explains, why magnetic friction
depends on $\alpha$ but not on $wS_\mathrm{tip}$ for low temperatures,
while it depends on $wS_\mathrm{tip}$ but not on $\alpha$ for high
temperatures where it decreases exponentially with $T$.   

Important extensions of the present investigation include the effects
of a tip magnetisation pointing in a different than the $z$-direction,
of the strength and sign of spin anisotropy, $d_z$, or of the
thickness of the magnetic layer.  Both, spin anisotropy and lattice
dimension will be crucial for the critical behaviour, as well as for
the critical temperature itself. Studies dealing with these quantities
are already in progress and will be reported in a future work.

\acknowledgements 
This work was supported by the German Research Foundation (DFG) via
SFB 616 ``Energy dissipation at surfaces''. Computation time was
granted in J\"ulich by the John-von-Neumann Institute of Computing
(NIC).

\bibliographystyle{eplbib}
\bibliography{paper}

\end{document}